%% file: main.tex
\documentclass[sigconf,authorversion,nonacm]{acmart}

\AtBeginDocument{%
  \providecommand\BibTeX{{%
    Bib\TeX}}}

\usepackage{graphicx}  
\usepackage{tabularx}   
\usepackage{makecell}   
\usepackage[most]{tcolorbox}
\usepackage{fontawesome5}
\usepackage{caption}

\usepackage{enumitem}

\setlist{topsep=2pt, itemsep=1pt, parsep=0pt, partopsep=0pt}

\tcbset{before skip=4pt, after skip=-3pt}
\begin{document}


\title[Empirical Investigation of Quantum Computing Toolchains and Algorithms]{Empirical Investigation of Quantum Computing Toolchains and Algorithms : Mining Stack Overflow Repository }

\author{Maryam Tavassoli Sabzevari}
\affiliation{%
  \institution{M3S Empirical Software Engineering Research Unit,\\ University of Oulu}
  \city{90014 Oulu}
  \country{Finland}}
\email{Maryam.TavassoliSabzevari@oulu.fi}

\author{Arif Ali Khan}
\affiliation{%
  \institution{M3S Empirical Software Engineering Research Unit,\\ University of Oulu}
  \city{90014 Oulu}
  \country{Finland}}
\email{Arif.Khan@oulu.fi}

\renewcommand{\shortauthors}{MT Sabzevari et al.}

\begin{abstract}
Quantum computing (QC) is increasingly transitioning toward practical and industrial adoption, highlighting the need to understand how developers engage with quantum technologies. In this study, we analyze 1,404 Stack Overflow posts related to quantum computing topics, including quantum programming, tools, and algorithms, to investigate real-world developer discussions. Using topic modeling and quantitative analysis, we identify the main discussion topics, their popularity, and the tools, programming languages, and quantum algorithms referenced by practitioners. We further assess the difficulty of developer questions using two metrics: (i) the percentage of questions without accepted answers and (ii) the median time required to receive an accepted answer. Our findings reveal seven main topics, with hybrid quantum--classical computing and quantum circuit implementation emerging as the most prevalent. We observe that Qiskit and Q\# dominate developer discussions, while Grover’s and Shor’s algorithms are the most frequently referenced. Moreover, our analysis highlights differences in engagement and difficulty across topics, tools, and algorithms, indicating varying levels of maturity and community support. These findings provide actionable insights for researchers, tool developers, and educators, supporting improvements in usability, documentation, and learning resources in quantum software engineering. To support transparency and reproducibility, the open-source dataset used in this study is publicly available at~\cite{tavassoli2026zenodo}.
\end{abstract}



\maketitle

\noindent\textit{This is the author version of a paper accepted at the FSE Companion 2026. The final version is available via the ACM Digital Library.}

\section{Introduction}
Quantum computing (QC) is rapidly transitioning from theoretical research to practical application, with increasing adoption across domains such as cybersecurity, robotics, finance, and healthcare. Its potential has also been demonstrated in emerging areas such as agriculture and environmental science~\cite{aithal2023advances}. This growing relevance has attracted substantial industrial interest, with leading technology firms such as IBM~\cite{IBM}, Google~\cite{Google}, and Microsoft~\cite{Microsoft} investing heavily in QC research and development.

The development of robust quantum software systems is essential to leverage quantum hardware and fully exploit the potential of quantum information processing~\cite{khan2023software}. However, quantum software development introduces unique complexities arising from quantum mechanical phenomena such as superposition and entanglement~\cite{horodecki2009quantum}. To address these challenges, the emerging discipline of quantum software engineering (QSE)~\cite{zhao2020quantum,piattini2021toward,piattini2020talavera} aims to adapt conventional software engineering practices for quantum application development. Despite its promise, QSE faces several barriers, including the inherent difficulty of quantum programming and the continued reliance on classical infrastructures for quantum circuit compilation, largely due to the lack of quantum data storage technologies~\cite{nguyen2024qfaas}.
To mitigate these challenges, an increasing number of platforms, frameworks, and programming languages have been introduced to facilitate access to quantum resources. While this diversity provides flexibility, it also creates practical challenges for developers, including difficulty in selecting appropriate tools, understanding common development obstacles, and effectively implementing quantum algorithms in real-world scenarios.
To address these gaps, this study investigates developer discussions in quantum computing through the following research questions:
\begin{itemize}
    \item RQ1. What topics do developers discuss in quantum computing?
    \item RQ2. What tools and platforms are used by practitioners to access quantum computing resources?
    \item RQ3. What quantum algorithms are implemented or referenced by practitioners?
    \item RQ4. How are tools and algorithms distributed across topics, and how do these distributions differ in terms of question difficulty?
\end{itemize}
First, we examine the topics developers most frequently discuss, aiming to understand the primary areas of concern and interest. Second, we analyze the tools and platforms used by practitioners and assess the difficulty of related questions, providing insights into the usability and support landscape of quantum development environments. Third, we investigate the quantum algorithms referenced by developers and evaluate the difficulty associated with their implementation and usage.
The main contributions of this study are as follows:
\begin{itemize}
    \item An empirical analysis of developer discussions on Stack Overflow, identifying the dominant topics in quantum computing and revealing key areas of practitioner interest.
    
    \item A quantitative evaluation of the tools and platforms used by practitioners, together with an assessment of question difficulty, highlighting differences in usability, maturity, and community support.
    
    \item An analysis of quantum algorithms referenced in practice, along with their associated difficulty, providing insights into implementation complexity and potential gaps in tooling and guidance

    \item A cross-analysis of tools, algorithms, and topics, highlighting how their distribution is associated with variations in question difficulty.
\end{itemize}
The remainder of this paper is organized as follows. Section~\ref{sec:Background} presents the background and motivation. Section~\ref{sec:method} describes the research methodology. Section~\ref{sec:result} reports the results and discussion. Section~\ref{sec:contribution} outlines the contributions and implications of the study. Section~\ref{sec:threats} discusses threats to validity, and Section~\ref{sec:conclusion} concludes the paper.

\section{Background and Motivation}
\label{sec:Background}
\input{Sections/BackgroundandMotivation}

\section{Methodology}
\label{sec:method}
\input{Sections/Methodology}

\section{Results and Discussion}
\label{sec:result}
\input{Sections/ResultAndDiscussion}

\section{Expected Contributions and Implications}
\label{sec:contribution}
\input{Sections/ExpectedContributions}

\section{Threats to Validity}
\label{sec:threats}
\input{Sections/ThreatsToValidity}

\section{Conclusion and Future Work}
\label{sec:conclusion}
\input{Sections/Conclusion}

\section{Acknowledgment}
\input{Sections/Acknowledgment}

\bibliographystyle{ACM-Reference-Format}
\bibliography{main}

\end{document}

%% file: Sections/BackgroundandMotivation.tex
Quantum computing is an emerging field in which the term \textit{quantum} refers to the principles of quantum mechanics that enable systems to perform computationally intensive operations~\cite{zhao2020quantum,nielsen2010quantum}. In physics, a quantum represents the smallest discrete unit of a physical property and typically refers to atomic or subatomic particles, such as electrons, neutrons, and photons. Quantum computing (QC) leverages these principles to execute certain computational tasks with higher efficiency compared to conventional computing systems. Unlike classical computers, which evaluate possibilities sequentially, quantum computers can explore multiple possibilities simultaneously. QC is particularly well-suited for problems such as optimization, simulation, cryptography, data analysis, and molecular modeling~\cite{zhao2020quantum,mcardle2020quantum,bova2021commercial}.

The fundamental difference between classical and quantum computers lies in their basic unit of information. Classical computers operate on \textit{bits}, which can represent either 0 or 1. In contrast, quantum computers utilize \textit{qubits}, which can exist in a superposition of both states simultaneously. Mathematically, a qubit is expressed as:
\begin{equation}
|\psi\rangle = x|0\rangle + y|1\rangle
\end{equation}
where $|\psi\rangle$ denotes the quantum state of the qubit, $|0\rangle$ and $|1\rangle$ are the computational basis states, and $x$ and $y$ are complex-valued probability amplitudes. The amplitudes $x$ and $y$ satisfy the normalization condition:
\begin{equation}
|x|^2 + |y|^2 = 1
\end{equation}
ensuring that the total probability of observing the qubit in either $|0\rangle$ or $|1\rangle$ upon measurement equals one.

Superposition is a defining feature of quantum mechanics and forms the foundation of quantum computational parallelism. While a classical register of $n$ bits can represent only one of $2^n$ possible states at a time, a quantum register with $n$ qubits can exist in a superposition of all $2^n$ states simultaneously, enabling parallel exploration of the state space.

Another key characteristic is \textit{entanglement}, a phenomenon in which two or more quantum systems become strongly correlated such that the state of one cannot be described independently of the others. Notably, entanglement is not constrained by physical distance: measurements on one entangled system can instantaneously influence the state of the other~\cite{zhao2020quantum}. Together, superposition and entanglement enable quantum computers to solve certain classes of problems more efficiently than their classical counterparts.

To facilitate access to quantum resources, various tools, platforms, and quantum programming languages have been introduced by both academia and industry \cite{singh2024survey}. 

Several studies have investigated the topics and challenges discussed and faced by developers when working with quantum technologies. For instance, Li et al.~\cite{li2021understanding} analyzed quantum software engineering (QSE) topics in technical forums. Similarly, Khan et al.~\cite{khan2025mining} conducted a mining study of question-and-answer platforms to examine quantum software programming, including tools, topics, and challenges. Husain et al.~\cite{husain2025exploring} further explored developer discussions to identify challenges in quantum software engineering.

Despite these efforts, existing studies typically focus on specific aspects such as quantum programming, QSE challenges, or individual tools and algorithms. To the best of our knowledge, there is no comprehensive study that jointly examines tools, platforms, programming languages, and quantum algorithms from the perspective of practitioners engaged in their practical use.

To address this gap, we conduct a mining study on Stack Overflow, analyzing discussions related to quantum computing, including tools, platforms, and quantum algorithms referenced by developers.

%% file: Sections/Methodology.tex
This study follows an empirical approach based on mining Stack Overflow data to investigate developer discussions in the context of quantum computing. The aim of this study is to analyze the dominant topics discussed in quantum computing-related posts on Stack Overflow, as well as to investigate the tools, platforms, and programming languages used by practitioners, and the quantum algorithms they reference. To achieve this, a six-step methodology was followed to answer the defined research questions.
\begin{itemize}
    \item RQ1. What topics do developers discuss in quantum computing? 
    \textit{Rationale:} Understanding developer-discussed topics helps identify key areas of interest and practical challenges in quantum computing, particularly given the early stage of quantum software engineering.    
    \item RQ2. What tools and platforms are used by practitioners to access quantum computing resources? 
    \begin{itemize}
        \item RQ2.1. How difficult are questions related to these tools and platforms?
    \end{itemize}
    \textit{Rationale:} Analyzing the tools and platforms used by developers, along with the difficulty of related questions, provides insights into their usability, maturity, and community support.    
    \item RQ3. What quantum algorithms are implemented or referenced by practitioners? 
    \begin{itemize}
        \item RQ3.1. How difficult are questions related to these algorithms?
    \end{itemize}
    \textit{Rationale:} Investigating discussed quantum algorithms and their associated difficulty reveals practical challenges in implementation and highlights gaps in tooling and guidance.
    \item • RQ4. How are tools and algorithms distributed across topics, and how do these distributions differ in terms of question difficulty?
    \textit{Rationale:} Analyzing how tools and algorithms are distributed across topics, and comparing their associated question difficulty, provides a more detailed understanding of variations in developer discussions and question resolution across quantum computing topics.
\end{itemize}
\subsection{Step 1: Downloading and Extracting the Stack Overflow Data Dump}

The complete Stack Overflow data dump (last update: 31 December 2025) \cite{stackexchange_archive_2025} was downloaded. This dataset contains questions, answers, and associated metadata (e.g., view count, creation date, and accepted answer status). The initial dataset includes approximately 60 million posts.

\subsection{Step 2: Identification of Quantum-Related Tags}

Stack Overflow contains posts covering a wide variety of software-related topics. Posts are typically tagged by their authors using commonly used tags to improve visibility and increase the likelihood of receiving answers \cite{barua2014developers}.

Following prior work \cite{abdellatif2020challenges}, a tag-based approach was adopted to identify posts relevant to quantum computing. First, all posts tagged with \texttt{quantum-computing}, \texttt{qiskit}, and \texttt{qubit} were extracted, resulting in 597 question posts (after deduplication) and 1258 records including answers. Subsequently, all tags co-occurring with these initial tags were extracted for further analysis.

Two heuristic measures were used to expand the set of quantum-related tags, following prior studies \cite{wan2019programmers, rosen2016mobile}.

\noindent\textbf{Tag Relevance Threshold (TRT).} This metric measures how strongly a tag is associated with quantum computing:

\begin{equation}
TRT_{tag} =
\frac{\#(posts\ in\ our\ dataset\ with\ that\ tag)}
{\#(all\ posts\ with\ that\ tag)}
\end{equation}

This metric helps eliminate irrelevant tags.

\noindent\textbf{Tag Significance Threshold (TST).} This metric measures the prominence of a tag within the quantum-related dataset:

\begin{equation}
TST_{tag} =
\frac{\#(posts\ in\ our\ dataset\ with\ that\ tag)}
{\#(posts\ in\ our\ dataset\ for\ the\ most\ frequent\ tag)}
\end{equation}

In our dataset, the most frequent tag is \texttt{quantum-computing} with 406 occurrences.

A discrepancy may exist between the number of posts retrieved for a given tag from the Posts dataset and the corresponding value reported in the \texttt{Count} attribute of \texttt{Tags.xml}. This is due to the exclusion of deleted posts from the Posts dataset, which may still be reflected in the tag counts.

To identify relevant tags, those exceeding predefined TRT and TST thresholds were selected. The first author examined candidate tags under different threshold values, and the selection was iteratively refined and discussed among the authors until consensus was reached, reducing potential bias. Alternative threshold values were explored and produced comparable tag sets.

Table~\ref{tab:top-tags} shows the first ten tags sorted by TST and TRT, respectively. It can be seen that the thresholds have been selected in such a way that although we have maintained the relevant posts but noisy ones will be eliminated.
\begin{table}[!b]
\centering
\small
\renewcommand{\arraystretch}{1.1}
\caption{Top 10 tags with TRT and TST metrics}
\label{tab:top-tags}
\begin{tabular}{lrrrr}
\toprule
Tag Name & Record Count & Total Posts & TRT & TST \\
\midrule
quantum computing & 406 & 435 & 93.33\% & 100.00\%\\
qiskit & 319 & 335 & 95.22\% & 78.57\% \\
python & 190 & 2222191 & 0.01\% & 46.80\%\\
q\# & 61 & 131 & 46.56\% & 15.02\% \\
python 3.x & 38 & 342762 & 0.01\% & 9.36\% \\
qubit & 27 & 28 & 96.43\% & 6.65\% \\
algorithm & 19 & 121985 & 0.02\% & 4.68\% \\
jupyter notebook & 17 & 25364 & 0.07\% & 4.19\% \\
pip & 17 & 24608 & 0.07\% & 4.19\% \\
machine learning & 15 & 56626 & 0.03\% & 3.69\% \\
\bottomrule
\end{tabular}
\end{table}
Based on this process, tags with TRT $> 40\%$ and TST $> 4\%$ were selected. These thresholds are consistent with prior studies using similar methodologies \cite{abdellatif2020challenges, ahmed2018concurrency, bagherzadeh2019going}. The final set of selected tags includes: \textbf{qubit, qiskit, quantum-computing, and Q\#}.
\subsection{Step 3: Extraction of RQ-Related Posts}
All posts containing at least one of the selected tags were extracted. After removing duplicates, the dataset includes 666 question posts and 738 answer posts, resulting in a total of 1404 posts.
In this study, the primary unit of analysis is the question post, with titles used for textual analysis and answers used for auxiliary metrics.
\subsection{Step 4: Text Preprocessing for Topic Modeling}
Post titles were used for topic modeling, as they typically summarize the main concept of a post, while the body may contain additional or unrelated information \cite{rosen2016mobile, abdellatif2020challenges}. This choice reduces noise and ensures consistency across posts, while capturing the main intent of each question.
Before automated preprocessing, a manual normalization step was performed to preserve domain-specific programming language identifiers. In particular, occurrences of language names containing the symbol \# were normalized to textual equivalents (e.g., Q\# $\rightarrow$ qsharp, C\# $\rightarrow$ csharp, and F\# $\rightarrow$ fsharp).

\begin{itemize}
    \item The normalized titles were preprocessed using Python’s NLTK library \cite{bird2009nltk}. Titles were converted to lowercase, and punctuation was removed while preserving underscores. Common English stopwords were removed using the NLTK stopword corpus \cite{sebleier2010stopwords}. Additionally, a set of generic programming terms (e.g., use, get, run, find, import, code, and error) was removed to reduce noise.
    
    \item A bigram model was constructed using the Gensim library \cite{gensim_website} to capture frequently co-occurring word pairs. Detected pairs were merged into single tokens (e.g., quantum computing $\rightarrow$ \texttt{quantum\_computing}).
    
    \item Lemmatization was applied using NLTK’s WordNetLemmatizer \cite{bird2009nltk}. Tokens were first assigned part-of-speech (POS) tags and then mapped to WordNet categories before lemmatization. Bigrams were preserved as single tokens.
\end{itemize}

A final cleanup step removed numeric-only tokens, generic tokens, and single-character tokens. All removed tokens were recorded in a log file for verification. During manual inspection, certain single-character tokens (e.g., u, x, v, z, q, and r) were identified as domain-relevant (e.g., quantum gates or notation) and were preserved through a whitelist mechanism. The resulting processed titles were used as input for topic modeling.

\subsection{Step 5: Topic Modeling}

To identify the main topics discussed in the collected Stack Overflow posts (RQ1), Latent Dirichlet Allocation (LDA) \cite{blei2003latent}, a widely used probabilistic topic modeling technique, was applied.

Each question title was treated as an individual document. After preprocessing, the dataset consisted of 666 cleaned titles. A dictionary was constructed, and frequency-based filtering was applied using Gensim’s \texttt{filter\_extremes} function. Tokens appearing in fewer than two documents or in more than 50\% of documents were removed, reducing the vocabulary size from 1275 to 483 tokens.

To determine the number of topics ($K$), a two-stage evaluation approach was adopted.

\textbf{Stage 1: Preliminary Screening.} Candidate values of $K$ ranging from 5 to 15 were evaluated using the \texttt{u\_mass} coherence metric to identify promising candidates.

\textbf{Stage 2: Coherence-Based Evaluation.} Shortlisted models ($K = 6, 7, 9$) were evaluated using the \texttt{c\_v} coherence metric, which correlates with human interpretability \cite{10.1145/2684822.2685324}. The model with $K = 7$ achieved the highest coherence score ($c_v = 0.3423$).

\textbf{Stage 3: Manual Topic Inspection.} Topics were manually inspected based on top words and representative question titles. The $K = 7$ model provided the best balance between interpretability and topic distinctiveness.

Based on both quantitative and qualitative evaluation, $K = 7$ was selected. Each document was assigned to its dominant topic based on the highest topic probability. Topic prevalence was then computed based on the distribution of documents across topics.

Finally, topics were manually labeled based on their most representative words and associated question titles.

\subsection{Step 6: Data Processing and Analysis}

This step operationalizes the analysis required to answer RQ1–RQ3 by defining the data processing and analytical procedures applied to the extracted dataset.

\subsubsection{Topic Popularity Analysis (RQ1)}

To investigate the popularity of the identified topics, Stack Overflow posts assigned to each topic were analyzed using two complementary metrics commonly adopted in prior studies: the average number of views and the average score of posts \cite{abdellatif2020challenges, ahmed2018concurrency, bagherzadeh2019going, nadi2016jumping}.

Question posts were grouped according to their assigned dominant topic. For each topic, the average number of views (\textit{avg. views}) was computed to capture the level of community interest, as higher view counts indicate that more developers accessed posts related to that topic. In addition, the average score (\textit{avg. score}) of posts within each topic was calculated. The score reflects the perceived usefulness of a post, as it is derived from user up-votes on Stack Overflow.

To improve the robustness of the analysis and account for potential skewness in the distribution of views and scores, median values for both metrics are also reported.

\subsubsection{Identification of Quantum Tools/Languages in Practitioner Discussions (RQ2)}

To answer RQ2, which investigates the tools, platforms, and programming languages used by practitioners to access quantum computing resources, a rule-based data processing and analysis pipeline was designed using Stack Overflow question titles.

A preprocessed dataset of Stack Overflow question titles related to quantum computing was used. Titles had previously undergone normalization to ensure consistency. In particular, programming language identifiers containing special characters (e.g., Q\#, C\#, F\#) were converted into textual forms (e.g., \textit{qsharp}, \textit{csharp}, \textit{fsharp}).

We constructed a candidate tool dictionary based on domain knowledge and inspection of Stack Overflow tags and relevant ecosystems. The dictionary was manually curated and encoded in a standard format.

The dictionary construction and refinement process was guided by both domain knowledge and empirical inspection of Stack Overflow data, and was validated through manual verification of extracted matches.

Each entry in the dictionary consists of:
\begin{itemize}
    \item a canonical name (e.g., \textit{qiskit}, \textit{qsharp}, \textit{ibm quantum}),
    \item a set of variants (e.g., \textit{ibmq}, \textit{ibm\_q}, \textit{ibm runtime}), and
    \item a type category (e.g., quantum tool, quantum platform, cloud platform, programming language).
\end{itemize}

Variants were sorted by length in descending order to prioritize more specific expressions and reduce partial matching issues.

A rule-based extraction approach using dictionary matching over normalized titles was applied. For each candidate, regular expression patterns were generated from its variants.

Matching was performed using boundary-aware regular expressions to avoid partial matches within larger tokens.

Each title was processed independently, and for each candidate, at most one matching variant was recorded. This design avoids duplicate matches but may not capture multiple variant forms of the same tool within a single title.

For each post, the list of matched canonical tools/platforms, the corresponding matched variants, and their associated type categories were extracted. Extraction was performed exclusively on question titles and did not include post bodies.

To improve extraction coverage, a subset of original (non-preprocessed) titles was manually inspected to identify missing lexical variants.

Based on this inspection, the dictionary was extended with additional variants related to specific ecosystems, including:
\begin{itemize}
    \item IBM Quantum (e.g., IBMQ, IBM runtime, IBM quantum provider),
    \item Microsoft Quantum Development Kit (e.g., Microsoft QDK, quantum development kit), and
    \item Q\# ecosystem variants (e.g., Microsoft qsharp .NET).
\end{itemize}

The extraction process was then re-executed using the refined dictionary. This refinement led to a modest increase in coverage (from 374 to 377 posts with detected tool mentions) and improved identification of platform-specific references.

Following extraction, a frequency analysis was conducted to quantify tool usage.

Since a single post may mention multiple tools, multi-tool entries were split into individual records before aggregation.

The number of posts mentioning each tool, the percentage relative to all posts, and the percentage relative to posts containing at least one detected tool were then computed.

\subsubsection{Identification of Quantum Algorithms in Practitioner Discussions (RQ3)}

To answer RQ3, a multi-stage extraction process was adopted to identify quantum algorithms explicitly referenced in Stack Overflow posts. The process was designed to balance precision and recall by combining title-based and body-based analysis.

We first constructed a candidate list of quantum algorithms based on well-established algorithms commonly discussed in the literature and practice. The list included canonical quantum algorithms (e.g., Grover, Shor, Deutsch, Deutsch--Jozsa), core subroutines (e.g., Quantum Fourier Transform, Phase Estimation), and NISQ algorithms (e.g., Variational Quantum Eigensolver (VQE), Quantum Approximate Optimization Algorithm (QAOA)), as well as selected advanced algorithms (e.g., HHL) and related techniques (e.g., Swap Test, Amplitude Amplification).

For each algorithm, a set of textual variants was defined to account for naming variations, and these variants were mapped to canonical names to ensure consistent aggregation. In the first extraction stage, preprocessed question titles were analyzed to identify explicit mentions of algorithm names using boundary-aware regular expression matching. For each post, the extracted information included matched algorithm names, textual variants, algorithm types, and the number of detected algorithms. This stage identified 39 posts (5.9\%) containing explicit algorithm mentions in titles.

Given the low coverage, a second extraction stage was applied to the bodies of posts without algorithm mentions in titles (627 posts). The same dictionary and matching strategy were reused, with lightweight normalization applied to the text. This stage identified 33 additional posts containing algorithm mentions. The results from both stages were then combined by merging matched algorithms per post and removing duplicates, resulting in a total of 72 posts (10.8\%) containing at least one algorithm mention. Finally, a frequency analysis was conducted on the combined dataset to assess the distribution of algorithm mentions.
\subsubsection{Difficulty Analysis of Questions (RQ2 and RQ3)}

To assess the difficulty of questions related to quantum computing tools and algorithms, two commonly used metrics from prior studies were adopted \cite{abdellatif2020challenges, bagherzadeh2019going, rosen2016mobile}: (i) the percentage of questions without accepted answers and (ii) the median time to receive an accepted answer.

\begin{equation}
\small
\%\,\text{Without Accepted Answer} = 
\frac{\#\,\text{questions with no accepted answer}}{\#\,\text{total questions}} \times 100
\end{equation}

\begin{equation}
\small
\text{Time to Accepted Answer} = 
\text{CreationDate}_{\text{accepted answer}} - \text{CreationDate}_{\text{question}}
\end{equation}

The resulting time differences were converted to hours, and the median value was reported.

We use a merged dataset consisting of both questions and answers. Each record is identified using the \texttt{PostTypeId} field, where \texttt{PostTypeId = 1} indicates a question and \texttt{PostTypeId = 2} indicates an answer.

For this analysis, computations are restricted to question posts, while answer posts are used only to retrieve accepted answer timestamps.

To answer RQ2 and RQ3, questions are divided into two groups:

\begin{itemize}
    \item \textbf{RQ2 (Tools):}
    \begin{itemize}
        \item Tool-related questions
        \item Baseline: Without any detected tool questions
    \end{itemize}
    
    \item \textbf{RQ3 (Algorithms):}
    \begin{itemize}
        \item Algorithm-related questions
        \item Baseline: Without any detected algorithm questions
    \end{itemize}
\end{itemize}

These groupings enable comparative analysis between questions that explicitly reference tools or algorithms and those that do not.
For each group, we report total number of questions, number and percentage of questions without accepted answers, number of questions with accepted answers, and median time to accepted answer (in hours) to analysis the difficulty of questions.

%% file: Sections/ResultAndDiscussion.tex
This section presents the results of our analysis of Stack Overflow posts and their corresponding answers to address our research questions.

\subsection{RQ1. Topics}

Table~\ref{tab:topicLabels} presents the seven topics identified using the LDA model, along with the number and percentage of questions associated with each topic. The most prevalent topic is hybrid classical--quantum computing (23.42\%), followed by quantum circuit implementation issues (17.27\%). In contrast, the topic of quantum algorithms and circuit design represents the least frequent category (7.36\%).
\begin{table}[htbp]
\caption{Topic labels and distribution}
\label{tab:topicLabels}
\centering
\small
\setlength{\tabcolsep}{4pt}
\begin{tabular}{p{0.08\columnwidth} p{0.52\columnwidth} p{0.14\columnwidth} p{0.14\columnwidth}}
\hline
Topic & Topic Label & \# Posts & Percentage \\
\hline
T0 & Quantum Algorithms and Circuit Design & 49 & 7.36\% \\
T1 & Installation and Environment Configuration Issues & 93 & 13.96\% \\
T2 & Quantum Circuit Implementation Issues & 115 & 17.27\% \\
T3 & Quantum State Analysis and Measurement & 85 & 12.76\% \\
T4 & Hybrid Quantum-Classical Computing & 156 & 23.42\% \\
T5 & Q\# Programming and Project Development & 75 & 11.26\% \\
T6 & Quantum Hardware and Backend Execution Issues & 93 & 13.96\% \\
\hline
\end{tabular}
\end{table}
The predominance of hybrid classical--quantum computing questions reveals that integrating quantum and classical components is a central concern among developers. This suggests that current toolchains still provide limited support for managing hybrid workflows and their associated implementation complexity. Similarly, the relatively high proportion of questions related to circuit implementation reflects ongoing difficulties in translating conceptual quantum operations into executable circuits.

In contrast, questions on quantum algorithms and circuit design are less frequent. This may imply that fewer developers engage directly with low-level algorithm design, or that such discussions occur in more specialized venues. Overall, the distribution of topics highlights that practical development and integration issues dominate discussions, rather than foundational or theoretical aspects of quantum computing. Table~\ref{tab:topicPopularity} summarizes the popularity of each topic based on average views and scores.
\begin{table*}[!htbp]
\centering
\caption{Topic popularity based on average views and scores}
\label{tab:topicPopularity}

\setlength{\tabcolsep}{5pt}

\begin{tabular}{c c c c c c c c}
\hline
Topic & ID & \#Posts & (\%) & Avg. Views & Avg. Score & Median Views & Median Score \\
\hline
T0 & 0 & 49  & 7.36  & 719.82  & 1.55 & 345.0 & 1.0 \\
T1 & 1 & 93  & 13.96 & 1298.19 & 2.08 & 310.0 & 1.0 \\
T2 & 2 & 115 & 17.27 & 997.65  & 0.66 & 317.0 & 0.0 \\
T3 & 3 & 85  & 12.76 & 617.75  & 1.40 & 225.0 & 1.0 \\
T4 & 4 & 156 & 23.42 & 860.50  & 1.01 & 290.5 & 1.0 \\
T5 & 5 & 75  & 11.26 & 751.13  & 1.23 & 326.0 & 0.0 \\
T6 & 6 & 93  & 13.96 & 649.08  & 1.01 & 313.0 & 1.0 \\
\hline
\end{tabular}
\end{table*}
Although hybrid quantum--classical computing is the most prevalent topic, it does not receive the highest average number of views. Instead, installation and environment configuration issues exhibit the highest average views and scores, suggesting that foundational setup challenges attract broader attention and engagement from the community. This indicates that early-stage barriers in adopting quantum computing tools remain a significant concern. Conversely, topics such as quantum state analysis and measurement receive lower average views, likely reflecting a more specialized audience or narrower applicability. Similarly, circuit implementation issues, despite being common, are associated with relatively low average scores, indicating that these questions are more difficult to answer or less likely to receive highly rated responses. Overall, the popularity analysis complements the topic distribution by highlighting not only what developers ask about, but also which topics attract greater attention and engagement within the community.
\begin{tcolorbox}[
colback=gray!12,
colframe=gray!45,
arc=3mm,
boxrule=0.8pt,
width=\columnwidth,
enhanced,
breakable,
title={\faKey\hspace{0.5em}Key Insights for RQ1},
fonttitle=\bfseries,
coltitle=black,
boxed title style={
    colback=gray!20,
    colframe=gray!45,
    arc=2mm
},
attach boxed title to top left={xshift=4mm,yshift=-2mm},
]
Seven main topics were identified in developer discussions. Hybrid quantum--classical computing and circuit implementation were the most prevalent, while installation and environment issues attracted the highest community attention. Overall, discussions were dominated by practical development and adoption concerns.
\end{tcolorbox}
\subsection{RQ2. Tools and platforms}
Table~\ref{tab:candidate_frequency} summarizes the frequency of tools and platforms mentioned in quantum computing-related posts. Qiskit is the most frequently referenced tool, appearing in 31.23\% of all posts (55.17\% of matched posts), followed by Q\# (13.81\%). Other tools, such as IBM Quantum, Amazon Braket, and Cirq, appear less frequently.
\begin{table}[!t]
\centering
\small
\caption{RQ2. Tool Frequency Table}
\label{tab:candidate_frequency}
\begin{tabularx}{\columnwidth}{|c|X|c|c|c|}
\hline
\textbf{Rank} & \textbf{Candidate Name} & \textbf{Freq.} & \textbf{\% All} & \textbf{\% Matched} \\
\hline
1 & qiskit & 208 & 31.23 & 55.17 \\
2 & qsharp & 92 & 13.81 & 24.4 \\
3 & python & 33 & 4.95 & 8.75 \\
4 & ibm quantum & 23 & 3.45 & 6.1 \\
5 & qdk & 14 & 2.1 & 3.71 \\
6 & amazon braket & 9 & 1.35 & 2.39 \\
7 & qutip & 9 & 1.35 & 2.39 \\
8 & tensorflow quantum & 7 & 1.05 & 1.86 \\
9 & cirq & 4 & 0.6 & 1.06 \\
10 & qbraid & 3 & 0.45 & 0.8 \\
11 & fsharp & 3 & 0.45 & 0.8 \\
12 & amazon web services & 2 & 0.3 & 0.53 \\
13 & haskell & 2 & 0.3 & 0.53 \\
14 & qristal sdk & 2 & 0.3 & 0.53 \\
15 & node.js & 1 & 0.15 & 0.27 \\
16 & microsoft liquid & 1 & 0.15 & 0.27 \\
17 & azure quantum & 1 & 0.15 & 0.27 \\
18 & prolog & 1 & 0.15 & 0.27 \\
\hline
\end{tabularx}
\end{table}
The dominance of Qiskit likely reflect its widespread adoption and active community support, which likely increases its visibility in developer discussions. Similarly, the presence of Q\# as the second most frequent tool reveals that Microsoft’s quantum ecosystem is also actively used by practitioners. In contrast, other tools and languages, such as Prolog or domain-specific frameworks, appear only marginally, indicating limited usage or discussion within the Stack Overflow community.
{\setlength{\textfloatsep}{6pt}
\setlength{\intextsep}{6pt}
\begin{table}[!tbp]
\centering
\caption{RQ2. Tools Difficulty Metrics Summary}
\label{tab:rq1_difficulty}
\renewcommand{\arraystretch}{1.2}
\setlength{\tabcolsep}{2pt}
\small
\begin{tabular}{|
>{\centering\arraybackslash}m{1.65cm}|
>{\centering\arraybackslash}m{1.25cm}|
>{\centering\arraybackslash}m{1.45cm}|
>{\centering\arraybackslash}m{1.45cm}|
>{\centering\arraybackslash}m{1.75cm}|}
\hline
\textbf{Group} &
\shortstack{\textbf{No. of}\\\textbf{questions}} &
\shortstack{\textbf{No. of}\\\textbf{accepted}\\\textbf{answers}} &
\shortstack{\textbf{\% no}\\\textbf{accepted}\\\textbf{answer}} &
\shortstack{\textbf{Median time}\\\textbf{to accepted}\\\textbf{answer (h)}} \\
\hline
\makecell{\textbf{tool-related}} & 377 & 225 & 59.68 & 8.47 \\[3pt]
\makecell{\textbf{baseline}\\\textbf{(non-tool)}} & 289 & 181 & 62.63 & 12.18 \\
\hline
\end{tabular}
\end{table}
}
To assess the difficulty of questions related to these tools, we operationalize question difficulty using two metrics: (1) the percentage of questions without an accepted answer, and (2) the median time required to receive an accepted answer.

Table~\ref{tab:rq1_difficulty} compares tool-related posts with a baseline of non-tool-related quantum computing posts.

The results show that tool-related questions have a slightly lower percentage of unanswered posts compared to the baseline (59.68\% vs. 62.63\%). In addition, the median time to receive an accepted answer is shorter for tool-related questions (8.47 hours vs. 12.18 hours). 

These findings highlight that, while questions about quantum computing tools remain challenging overall, they are not substantially more difficult than other quantum computing questions. The relatively faster response time indicate that tool-related issues benefit from stronger community knowledge and support.

We note that our difficulty metrics are based on two indicators. In particular, the presence of an accepted answer does not necessarily guarantee correctness, and the time to receive an accepted answer may not fully capture the complexity of a question.

A more fine-grained analysis of difficulty across individual tools (e.g., Qiskit vs. Q\#) could provide deeper insights; however, this is left for future work.

The dominance of Qiskit may not only reflect its popularity, but also its role as a primary gateway to cloud-based quantum computing resources, particularly through IBM Quantum services. This implies that accessibility and integration with cloud infrastructures play a crucial role in shaping developer engagement. In contrast, tools with more limited ecosystems or steeper learning curves may receive less attention despite their technical capabilities.
\begin{tcolorbox}[
colback=gray!12,
colframe=gray!45,
arc=3mm,
boxrule=0.8pt,
width=\columnwidth,
enhanced,
breakable,
title={\faKey\hspace{0.5em}Key Insights for RQ2},
fonttitle=\bfseries,
coltitle=black,
boxed title style={
    colback=gray!20,
    colframe=gray!45,
    arc=2mm
},
attach boxed title to top left={xshift=4mm,yshift=-2mm},
]
Qiskit and Q\# were the most frequently discussed tools. Tool-related questions were resolved slightly faster and had marginally fewer unanswered posts than non-tool-related ones. However, support and difficulty varied substantially across tools.
\end{tcolorbox}
\subsection{RQ3. Quantum algorithms}

Table~\ref{tab:algorithm_frequency_combined_step3} presents the frequency of quantum algorithms discussed in Stack Overflow posts. Grover’s algorithm is the most frequently referenced (31.94\% of matched posts), followed by Shor’s algorithm (15.28\%). Other algorithms, such as the Quantum Fourier Transform, VQE, and QAOA, appear with moderate frequency, while algorithms such as phase estimation and Deutsch--Jozsa are less commonly discussed.

The prominence of Grover’s and Shor’s algorithms likely reflect their role as standard examples in quantum computing, which are widely used in introductory learning materials and educational resources. This increases their visibility and frequency in developer discussions. In contrast, more specialized or advanced algorithms appear less frequently, suggesting either a narrower audience or less practical engagement in typical development practice.

\begin{table}[!t]
\centering
\small
\caption{RQ3. Algorithm Frequency Table}
\label{tab:algorithm_frequency_combined_step3}
\begin{tabularx}{\columnwidth}{|c|X|c|c|c|}
\hline
\textbf{Rank} & \textbf{Algorithm} & \textbf{Count} & \textbf{\% All} & \textbf{\% Matched} \\
\hline
1 & grover & 23 & 3.45 & 31.94 \\
2 & shor & 11 & 1.65 & 15.28 \\
3 & quantum fourier transform & 10 & 1.5 & 13.89 \\
4 & vqe & 10 & 1.5 & 13.89 \\
5 & qaoa & 9 & 1.35 & 12.5 \\
6 & deutsch & 5 & 0.75 & 6.94 \\
7 & hhl & 5 & 0.75 & 6.94 \\
8 & deutsch-jozsa & 2 & 0.3 & 2.78 \\
9 & phase estimation & 2 & 0.3 & 2.78 \\
\hline
\end{tabularx}
\end{table}

To assess the difficulty of algorithm-related questions, we use the same metrics defined earlier: the percentage of questions without an accepted answer and the median time to receive an accepted answer.

Table~\ref{tab:algorithm_difficulty} compares algorithm-related questions with a baseline of non-algorithm-related quantum computing posts.
{\setlength{\textfloatsep}{6pt}
\setlength{\intextsep}{6pt}
\begin{table}[!tbp]
\centering
\caption{RQ3. Algorithm Difficulty Metrics Summary}
\label{tab:algorithm_difficulty}
\renewcommand{\arraystretch}{1.2}
\setlength{\tabcolsep}{2pt}
\small
\begin{tabular}{|
>{\centering\arraybackslash}m{1.65cm}|
>{\centering\arraybackslash}m{1.25cm}|
>{\centering\arraybackslash}m{1.45cm}|
>{\centering\arraybackslash}m{1.45cm}|
>{\centering\arraybackslash}m{1.75cm}|}
\hline
\textbf{Group} &
\shortstack{\textbf{No. of}\\\textbf{questions}} &
\shortstack{\textbf{No. of}\\\textbf{accepted}\\\textbf{answers}} &
\shortstack{\textbf{\% no}\\\textbf{accepted}\\\textbf{answer}} &
\shortstack{\textbf{Median time}\\\textbf{to accepted}\\\textbf{answer (h)}} \\
\hline
\makecell{\textbf{algorithm}\\[2pt]\textbf{related}} & 72 & 50 & 69.44 & 6.88 \\[13pt]
\makecell{\textbf{baseline}\\[2pt](non-algo)} & 594 & 356 & 59.93 & 10.47 \\
\hline
\end{tabular}
\end{table}
}
The results indicate that algorithm-related questions have a higher proportion of unanswered posts compared to the baseline (69.44\% vs. 59.93\%), suggesting that they are relatively more difficult to resolve. 

However, the median time to receive an accepted answer is shorter for algorithm-related questions (6.88 hours vs. 10.47 hours). This indicate that, although such questions are more challenging overall, those that receive answers tend to be addressed relatively quickly, possibly by contributors familiar with these standard algorithms.

Overall, these findings highlight that algorithm-related discussions involve higher conceptual complexity, while still benefiting from expertise within the community.

An interesting observation is the discrepancy between the high frequency of tool-related discussions and the relatively low frequency of algorithm-related discussions. This suggests that developers may primarily engage with quantum computing at the level of tool usage and implementation, rather than at the level of algorithm design. This gap highlights a potential need for better abstractions, documentation, and educational resources that bridge the conceptual understanding of quantum algorithms with their practical implementation in existing toolchains.

One possible explanation is that many developers rely on pre-implemented algorithmic components provided by existing frameworks, reducing the need to explicitly reason about or implement algorithms from first principles. This further emphasizes the abstraction gap between high-level algorithmic concepts and their practical usage in current quantum development environments.

By combining the results across RQ1--RQ3, we observe that topics related to hybrid computing and circuit implementation (RQ1) are strongly aligned with the dominance of tool-related discussions (RQ2), while algorithm-related discussions remain relatively limited (RQ3). This indicates that current developer activity is driven more by practical implementation challenges within existing toolchains rather than by algorithmic innovation. This reinforces the need for improved abstractions and developer support that bridge high-level quantum algorithms with their implementation in concrete platforms.
\begin{tcolorbox}[
colback=gray!12,
colframe=gray!45,
arc=3mm,
boxrule=0.8pt,
width=\columnwidth,
enhanced,
breakable,
title={\faKey\hspace{0.5em}Key Insights for RQ3},
fonttitle=\bfseries,
coltitle=black,
boxed title style={
    colback=gray!20,
    colframe=gray!45,
    arc=2mm
},
attach boxed title to top left={xshift=4mm,yshift=-2mm},
]
Grover’s and Shor’s algorithms were the most frequently referenced. Algorithm-related questions had a higher proportion of unanswered posts, indicating greater difficulty, although resolved questions received accepted answers relatively quickly. Overall, algorithm discussions were less frequent than tool-related discussions.
\end{tcolorbox}
\subsection{RQ4. Tools, Algorithms, and Topics}
To further understand the interplay between tools, algorithms, and topics, we conducted a cross-analysis that examines (i) the distribution of tools and algorithms across topics, and (ii) their associated difficulty levels. Difficulty is measured using two metrics: (1) the percentage of questions without accepted answers, and (2) the median time to receive an accepted answer (in hours).

\begin{figure*}[!t]    
    \centering
    \includegraphics[width=\textwidth]{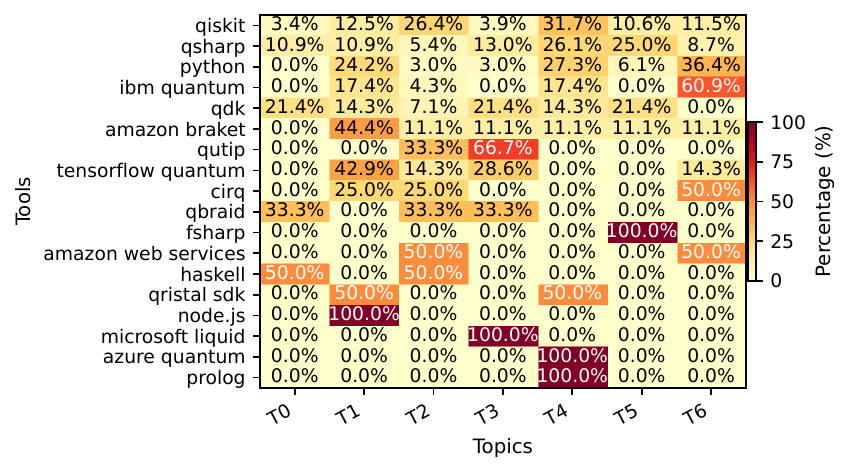}
    \caption{Tool–Topic Percentage}
    \label{fig:ToolTopic Percentage}
\end{figure*}

Figure~\ref{fig:ToolTopic Percentage} illustrates the distribution of tools across topics. Notably, Qiskit is highly concentrated in Hybrid Quantum–Classical Computing, indicating its central role in this domain. When considering difficulty (Table~\ref{tab:tool_difficulty}), Qiskit exhibits a relatively low median time to accepted answers (7.38 hours), suggesting an active and responsive community. This combination of high usage and relatively lower resolution time indicates that Qiskit may offer better support for developers entering hybrid quantum-classical development.

\begin{figure*}[!t]    
    \centering
    \includegraphics[width=\textwidth]{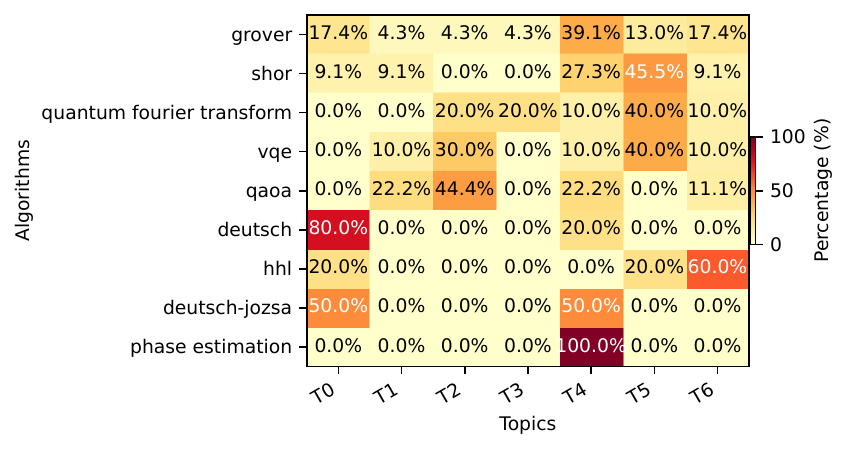}
    \caption{Algorithm–Topic Percentage}
    \label{fig:AlgorithmTopic Percentage}
\end{figure*}

Table~\ref{tab:tool_difficulty} also reveals substantial variability across tools. For example, QuTiP shows no accepted answers in the observed dataset, while Amazon Braket exhibits a high percentage of unresolved questions (77.78\%) and long response times. These findings suggest disparities in community support and maturity across quantum development platforms.
\begin{table}[!t]
\centering
\small
\renewcommand{\arraystretch}{1.25}
\setlength{\tabcolsep}{4pt}
\caption{Tool Difficulty Summary}
\label{tab:tool_difficulty}
\begin{tabular}{|l|c|c|c|c|}
\hline
\textbf{Tool} &
\textbf{Total} &
\makecell{\textbf{Without}\\\textbf{Accepted}} &
\makecell{\textbf{Pct Without}\\\textbf{Accepted}} &
\makecell{\textbf{Median Time}\\\textbf{(Hours)}} \\
\hline
qutip & 9 & 9 & 100.0\% & N/A \\
amazon braket & 9 & 7 & 77.7778\% & 188.657 \\
python & 33 & 23 & 69.697\% & 16.2258 \\
qiskit & 208 & 137 & 65.8654\% & 7.3863 \\
ibm quantum & 23 & 15 & 65.2174\% & 1.8368 \\
tensorflow quantum & 7 & 4 & 57.1429\% & 34.7708 \\
qdk & 14 & 6 & 42.8571\% & 35.512 \\
qsharp & 92 & 33 & 35.8696\% & 7.5371 \\
\hline
\end{tabular}
\end{table}
Figure~\ref{fig:AlgorithmTopic Percentage} presents the distribution of algorithms across topics. Deutsch’s algorithm is dominant in Quantum Algorithms and Circuit Design, while Shor, Quantum Fourier Transform (QFT), and VQE are prevalent in Q\# Programming and Project Development. However, these algorithms are associated with high difficulty levels, with at least 60\% of questions lacking accepted answers (Table~\ref{tab:Per algorithm difficulty summary}).
Among them, QFT stands out as particularly challenging, exhibiting both the highest percentage of unresolved questions (90\%) and the longest median time to accepted answers (78.94 hours). This suggests that despite its foundational role, QFT remains difficult to implement and troubleshoot in practice. These findings highlight opportunities for improving tooling, documentation, and educational resources around core quantum algorithms.

\begin{figure*}[!t]    
    \centering
    \includegraphics[width=\textwidth]{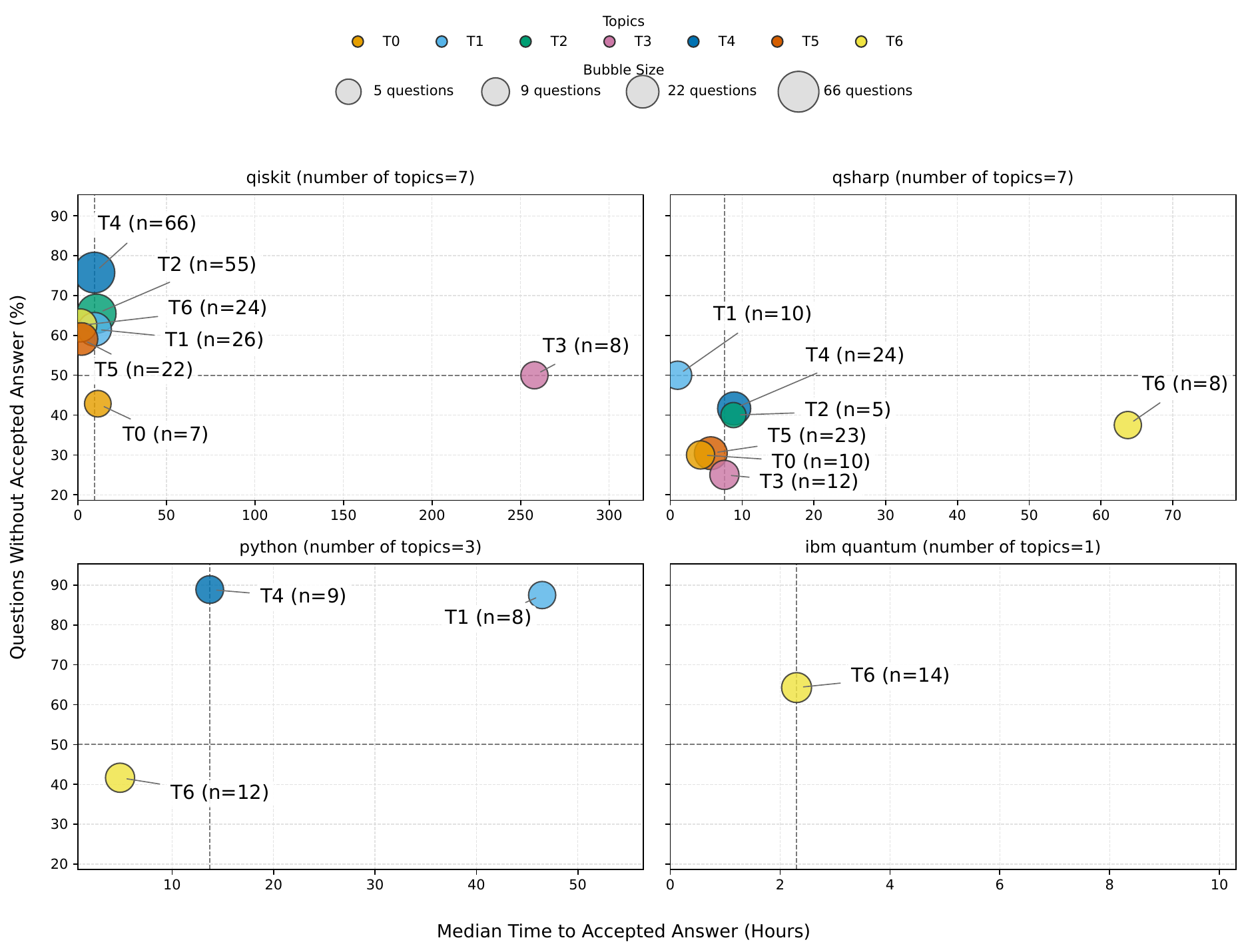}
    \caption{Tool topic difficulty}
    \label{fig:Tool topic difficulty}
\end{figure*}

{\setlength{\textfloatsep}{6pt}
\setlength{\intextsep}{6pt}
\begin{table}[!tbp]
\centering
\caption{Algorithm Difficulty Summary}
\label{tab:Per algorithm difficulty summary}
\renewcommand{\arraystretch}{1.15}
\setlength{\tabcolsep}{2pt}
\small
\begin{tabular}{|
>{\raggedright\arraybackslash}m{1.8cm}|
>{\centering\arraybackslash}m{0.65cm}|
>{\centering\arraybackslash}m{1.1cm}|
>{\centering\arraybackslash}m{1.6cm}|  
>{\centering\arraybackslash}m{1.05cm}|}
\hline
\textbf{Algorithm} &
\textbf{Total} &
\makecell{\textbf{Without}\\\textbf{Accepted}} &
\makecell{\textbf{Pct. Without}\\\textbf{Accepted}} &
\makecell{\textbf{Median}\\\textbf{Time (h)}} \\
\hline
QFT      & 10 & 9  & 90.0\%    & 78.9391 \\
QAOA     & 9  & 7  & 77.7778\% & 3.0907 \\
Shor     & 11 & 8  & 72.7273\% & 0.9142 \\
Grover   & 23 & 16 & 69.5652\% & 13.7448 \\
HHL      & 5  & 3  & 60.0\%    & 27.4284 \\
VQE      & 10 & 6  & 60.0\%    & 0.9131 \\
Deutsch  & 5  & 2  & 40.0\%    & 22.1156 \\
\hline
\end{tabular}
\end{table}
}
Finally, we performed a combined tool–topic difficulty analysis (Figure~\ref{fig:Tool topic difficulty}). To ensure robustness, we applied a minimum support threshold of five questions per tool–topic pair and excluded cases without valid accepted answers from time-based analysis.
The results reveal that difficulty is not uniform across topics for a given tool. For instance, Python shows significantly higher resolution times in Installation and Environment Configuration Issues compared to other topics. Similarly, Qiskit exhibits increased difficulty in Quantum Circuit Implementation Issues, while Q\# shows higher difficulty in Quantum Hardware and Backend Execution Issues. These discrepancies suggest that complexity is strongly context-dependent and not solely determined by the tool itself.
\begin{tcolorbox}[
colback=gray!12,
colframe=gray!45,
arc=3mm,
boxrule=0.8pt,
width=\columnwidth,
enhanced,
breakable,
title={\faKey\hspace{0.5em}Key Insights for RQ4},
fonttitle=\bfseries,
coltitle=black,
boxed title style={
    colback=gray!20,
    colframe=gray!45,
    arc=2mm
},
attach boxed title to top left={xshift=4mm,yshift=-2mm},
]
Tools and algorithms are unevenly distributed across topics. Qiskit is concentrated in hybrid quantum–classical computing and shows relatively faster resolution times, while other tools exhibit higher unresolved rates.

Algorithm usage is also topic-specific, with Deutsch dominant in algorithm-focused discussions and Shor, QFT, and VQE more common in Q\# development; these are generally associated with higher difficulty.

Overall, question difficulty varies across topics for both tools and algorithms, indicating that it is strongly context-dependent.
\end{tcolorbox}

%% file: Sections/ExpectedContributions.tex
This study provides insights into how practitioners engage with quantum computing on Stack Overflow. By analyzing developer discussions, the findings contribute to both research and practice in quantum software engineering.

First, the results related to topic distribution highlight that practical development concerns dominate practitioner discussions. In particular, topics such as hybrid quantum--classical computing and circuit implementation are more prevalent than foundational aspects such as quantum algorithm design. This suggests that current challenges primarily lie in integrating quantum and classical components and translating abstract quantum concepts into executable implementations. These findings can inform researchers in designing abstractions, frameworks, and tools that better support hybrid workflows.

Second, the popularity analysis reveals that early-stage adoption barriers remain significant. Installation and environment configuration issues exhibit the highest average views and scores, indicating strong community attention and demand for support. This suggests that improving toolchain usability, documentation, and onboarding processes should be a priority for both researchers and platform developers.

Third, the findings provide insights into the ecosystem of quantum computing tools and platforms. The dominance of Qiskit and Q\# (accounting for nearly 80\% of matched posts) indicates that a limited number of toolchains shape practitioner activity. At the same time, the relatively faster response time for tool-related questions suggests that these ecosystems benefit from active community support. These findings may serve as an indication for practitioners when selecting tools, particularly in contexts where community support and responsiveness are important considerations.

Fourth, the study highlights differences in the nature of algorithm-related discussions. Although fewer questions focus on quantum algorithms, they exhibit a higher proportion of unanswered posts, indicating increased conceptual difficulty. However, the shorter time to receive accepted answers suggests that expertise is available within the community. This implies that while common algorithms are well supported, educational resources and tooling for more advanced or less frequently used algorithms may still be limited.

Finally, this study contributes a comprehensive, practitioner-centered perspective on quantum computing by jointly analyzing topics, tools, and algorithms. These findings can support future research in quantum software engineering, including the design of developer-centric tools, improved documentation strategies, and more effective learning resources for quantum computing practitioners.

%% file: Sections/ThreatsToValidity.tex
According to Wohlin et al. \cite{Wohlin2024}, threats to validity may arise at different stages of the research process. 

Internal validity concerns the presence of additional factors that may influence the observed relationships. When investigating whether one factor affects another, there is a risk that the observed effect is influenced by an unobserved third variable. Lack of awareness of such factors, or their extent of influence, may introduce threats to internal validity. In this study, the initial dataset was constructed based on tag seeds, which may suffer from mislabeling (i.e., incorrectly assigned or missing tags) and could lead to the inclusion of irrelevant data or the omission of relevant data. To mitigate this threat, we employed the Tag Similarity Threshold (TST) and Tag Relevance Threshold (TRT), which have been used in prior studies to improve the coverage of topic-related posts \cite{abdellatif2020challenges, rosen2016mobile, wan2019programmers}. Furthermore, the selected threshold values are consistent with those reported in studies employing similar methodologies across different domains \cite{abdellatif2020challenges, ahmed2018concurrency, bagherzadeh2019going}. 

Another potential threat arises from relying solely on question titles for analysis. Although this approach may omit contextual details available in full question bodies, it has been widely adopted in prior research \cite{rosen2016mobile, abdellatif2020challenges}. 

Additionally, the selection of $K = 7$ for LDA topic modeling may affect the quality of the extracted topics, as $K$ determines the number of topics, and selecting an optimal value is inherently challenging. To mitigate this threat, we adopted a three-stage procedure consisting of preliminary screening, coherence-based evaluation, and manual inspection of topics based on top words and representative question titles.

Construct validity refers to the extent to which the operational measures accurately capture the intended concepts under investigation. In this study, we analyzed the difficulty and popularity of questions related to tools and algorithms using multiple metrics that have been validated and widely used in prior work \cite{abdellatif2020challenges, ahmed2018concurrency, bagherzadeh2019going, nadi2016jumping, rosen2016mobile}. 

External validity concerns the generalizability of the findings. In this study, the analysis is based solely on data from Stack Overflow. While other platforms may also host discussions on quantum computing, Stack Overflow is one of the most widely used forums among developers, which supports an acceptable level of generalizability. Nevertheless, the findings could be further strengthened by incorporating data from additional platforms or by conducting empirical studies involving practitioners working with quantum computing technologies.

%% file: Sections/Conclusion.tex
In this study, we analyzed Stack Overflow posts to investigate the topics, tools, and quantum algorithms discussed by developers in quantum computing. Our findings reveal seven distinct topics, with Hybrid Quantum-Classical Computing emerging as the most frequently discussed. In contrast, installation and environment configuration issues attract the highest number of views, indicating strong practical demand. Although quantum algorithms and circuit design are discussed less frequently, their higher average views and scores suggest deeper engagement and complexity.

With respect to tools, Qiskit is the most actively discussed framework, followed by Q\#. At the algorithm level, Grover’s and Shor’s algorithms dominate developer discussions. Interestingly, while the proportion of unanswered questions is only slightly lower than the baseline (approximately 3\% for tools and 10\% for algorithms), the median time to receive an accepted answer is shorter, indicating relatively efficient community support.
Finally, our cross-analysis shows that the difficulty of questions varies across topics for different tools and algorithms, indicating that question difficulty is context-dependent rather than solely determined by the tool or algorithm.

These findings highlight key challenges and engagement patterns in quantum software development, particularly in hybrid computing, tool usability, and algorithm implementation. They provide actionable insights for researchers and tool developers to improve documentation, usability, and support for quantum development environments.

As future work, we plan to extend this study by incorporating discussions from additional platforms and performing cross-platform comparisons of tool and algorithm difficulty.

%% file: Sections/Acknowledgment.tex
The author acknowledges the financial support provided by FAST, the Finnish Software Engineering Doctoral Research Network (project number: 2465000811), funded by the Ministry of Education and Culture, Finland. The author also acknowledges the use of large language models as writing aids to enhance the clarity and linguistic quality of this manuscript.